\documentclass[10pt]{article}
\pdfoutput=1
\usepackage{jheppub,amsmath,amssymb}
\usepackage{enumerate}
\usepackage{bm}
\usepackage{bbm}
\usepackage{enumitem}
\usepackage[usenames,dvipsnames]{xcolor}
\usepackage{amsmath}
\usepackage{amsfonts}
\usepackage{amssymb}
\usepackage{graphicx}   
\usepackage{hhline}
\usepackage{subfig}
\usepackage{hyperref}
\usepackage{color}
\usepackage{verbatim}
\usepackage{amsmath,amsthm,amssymb}
\usepackage{lscape}
\usepackage{float}
\usepackage{caption}
\usepackage{lscape}
\usepackage{breqn}
\usepackage{multicol}
\usepackage{graphics}
\usepackage{tikz,todonotes}
\usepackage[utf8]{inputenc}
\usepackage{autobreak}
\usepackage{verbatim}
\usepackage{mathtools}
\usepackage{tikz-feynman}
\usetikzlibrary{arrows,positioning,decorations.markings,decorations.pathmorphing,calc}
\pgfdeclarelayer{edgelayer}
\pgfdeclarelayer{nodelayer}
\usetikzlibrary{decorations.pathreplacing}
\pgfsetlayers{edgelayer,nodelayer,main}
\tikzset{none/.style={draw=none}}
\tikzset{new edge style 2/.style={black}}
\tikzset{new style 0/.style={black}}
\tikzset{rednode/.style={draw=none, scale=0.3pt,fill=red,circle, draw}}
\tikzset{redline/.style={line width=0.3mm,red}}
\tikzset{greyE/.style={line width=0.1mm,gray}}
\usepackage[utf8]{inputenc}
\usetikzlibrary{arrows,positioning,decorations.markings,decorations.pathmorphing,calc}

\usepackage{tensor}
\usepackage{longtable}

\definecolor{hyperref}{RGB}{026,028,087}

\newcommand{\beq}{\begin{equation}}
	\newcommand{\eeq}{\end{equation}}
\newcommand{\bea}{\begin{eqnarray}}
	\newcommand{\eea}{\end{eqnarray}}
\def\be{\begin{equation}}
	\def\ee{\end{equation}}

\def\beq{\begin{equation}}
	\def\eeq{\end{equation}}

\renewcommand{\L}{\mathcal L}

\def\be{\begin{equation}}
	\def\ee{\end{equation}}
\def\ba{\begin{eqnarray}}
	\def\ea{\end{eqnarray}}

\def\nn{\nonumber}

\usepackage[normalem]{ulem}

\newcommand{\mcg}[1]{\textcolor{blue}{[#1]}}

\def\ba{\begin{eqnarray}}
	\def\ea{\end{eqnarray}}

\def\L{\mathcal{L}}

\def\({\left(}
\def\){\right)}

\allowdisplaybreaks
\begin{document}

\preprint{Imperial/TP/2021/MC/04}	
	
	\title{Massive Double Copy in the High-Energy Limit}

	\author[a]{Mariana Carrillo González,}
	\author[a]{Arshia Momeni ,}
	\author[a]{Justinas Rumbutis}
	
	\affiliation[a]{Theoretical Physics, Blackett Laboratory, Imperial College, London, SW7 2AZ, U.K.}

	\emailAdd{m.carrillo-gonzalez@imperial.ac.uk}
	\emailAdd{arshia.momeni17@imperial.ac.uk}
	\emailAdd{j.rumbutis18@imperial.ac.uk}

	\abstract{The exploration of the massive double copy is still in its infancy, and only a few examples in limited contexts are known. In this paper, we analyze the newly discovered double copy for topologically massive theories beyond tree-level amplitudes in the high-energy (eikonal) limit. We find that contrary to the simple double copy relation that occurs in the massless $d\geq4$ case, the massive double copy requires information outside the eikonal limit to give correct results. However, a simple double copy relation arises for the phase shift. Additionally, we relate the eikonal amplitudes to the corresponding shockwave backgrounds and find a classical double copy relation that is manifest only for the appropriate choice of boundary conditions. This analysis hints towards the existence of a topologically massive double copy at all loop orders.}

	\maketitle
	


\section{Introduction}
The double copy is a surprising relationship that allows us to write gravitational theories as the ``square'' of gauge theories \cite{Kawai:1985xq,Bern:2008qj, Bern:2010ue}. While its original form consists of a relation between scattering amplitudes, it also applies more broadly to other observables and classical solutions \cite{Saotome:2012vy, Monteiro:2014cda, Luna:2015paa, Luna:2016due, White:2016jzc, Cardoso:2016amd, Luna:2016hge, Goldberger:2017frp, Ridgway:2015fdl, De_Smet_2017, Bahjat_Abbas_2017, Carrillo-Gonzalez:2017iyj,Anastasiou:2014qba,Anastasiou:2016csv,Anastasiou:2017taf,Anastasiou:2017nsz,Anastasiou:2018rdx,Borsten:2015pla,Goldberger_2018, Li_2018, Lee_2018, Plefka_2019, Berman_2019, Kim:2019jwm, Goldberger:2019xef, Alawadhi:2019urr, Banerjee:2019saj, CarrilloGonzalez:2019gof,Shen:2018ebu,Chacon:2021wbr,Bahjat-Abbas:2020cyb,Alfonsi:2020lub,Luna:2020adi,White:2020sfn,Alkac:2021bav,Alawadhi:2020jrv,Huang:2019cja,Keeler:2020rcv,Elor:2020nqe,Farnsworth:2021wvs,Lescano:2021ooe,Ferrero:2020vww,Gumus:2020hbb,Guevara:2021yud,Pasterski:2020pdk,Chacon:2021hfe,Alawadhi:2021uie,Cho:2021nim}. Recently, a new direction in the exploration of the applicability of the double copy has arisen. This consists of understanding the massive gauge theories which can satisfy colour-kinematics duality and lead to a physical double copy. In four dimensions, it has been noticed that the double copy of massive Yang-Mills amplitudes corresponds to dRGT massive gravity \cite{deRham:2010kj} (with a special choice of Wilsonian coefficients) amplitudes at 4-points \cite{Momeni:2020vvr,Johnson:2020pny}. Nevertheless, the 5-point double copy suffers from the appearance of non-physical (spurious) poles \cite{Johnson:2020pny}. The origin of these spurious poles is well understood. In the massive double copy, one can always construct kinematic numerators that satisfy the colour-kinematics duality, but this comes at the expense of having a more involved expression for the double copy in terms of the unshifted numerators. For example, at 4-points the double copy is
\begin{equation}
-i\left(\frac{\kappa}{2}\right)^{-2} M_{4}=\frac{n_{s}^{2}}{s-m^{2}}+\frac{n_{t}^{2}}{t-m^{2}}+\frac{n_{u}^{2}}{u-m^{2}}-\frac{\left(n_{s}+n_{t}+n_{u}\right)^{2}}{m^{2}} \ ,
\end{equation}
where $n_i$ are the unshifted kinematic numerators, i.e. $n_s+n_t+n_u\neq0 $. The amplitude above was obtained from the BCJ double copy of the massive Yang-Mills amplitude taking $c\rightarrow\tilde{n}$, where $\tilde{n}$ are the shifted numerators that satisfy $\tilde{n}_s+\tilde{n}_t+\tilde{n}_u=0 $. The generalization to 5-points leads to an amplitude with spurious poles. This can be avoided if one considers a theory with a tower of massive gauge bosons satisfying a special relationship between their masses which has been dubbed the spectral condition. Up till now, the only theory that has been shown to satisfy this condition for all its scattering amplitudes up to 5-points is the Kaluza-Klein theory where one compactifies the fifth dimension over $S^1$\cite{Momeni:2020hmc,Hang:2021fmp}. \\

In \cite{Gonzalez:2021bes}, a new possibility to avoid the spurious poles without requiring a tower of massive gluons was explored. By noticing a special feature of the massive double copy in 3 dimensions, it was shown that one can construct a physical double copy up to 5-points as long as the Yang-Mills amplitudes satisfy a single (instead of 4 as in the massless case) BCJ relation. Furthermore, it was shown that Topologically Massive Yang-Mills (TMYM) amplitudes satisfy such relation and that their double copy is the Topologically Massive Gravity (TMG) amplitudes \cite{Gonzalez:2021bes,Hang:2021oso}. An interesting question is whether the double copy relationship holds when we include matter interactions. This has been explored in different contexts in \cite{Moynihan:2020ejh,Burger:2021wss,Moynihan:2021rwh}, where it was shown that the double copy holds in a non-trivial manner and depends on the type of matter that scatters through topologically massive mediators. These studies suggest that the double copy holds for sources whose stress-energy tensor is traceless, otherwise a non-trivial relation arises.\\

In this paper, we will take a step forward in understanding the topologically massive double copy involving matter fields. First, we introduce the topologically massive theories including a minimal coupling to matter fields in Section \ref{sec:TMT}. We take a look at the 2-2 scattering of scalars through a massive mediator and find that the double copy requires an extra contact term interaction between the scalars. This extra term becomes subdominant in the eikonal limit in which the sources are highly energetic and their stress-energy tensor becomes traceless, leading to the standard double copy relation as suggested in 
\cite{Moynihan:2021rwh}. Given this, we will explore the eikonal limit in more detail in Section \ref{sec:eikonal}. We take advantage of the fact that both abelian and non-abelian objects can double copy to the same gravitational object \cite{Bahjat-Abbas:2020cyb} and look at the linearized TMYM case, that is, Topologically Massive Electrodynamics (TME). We prove that the TMG and TME amplitudes exponentiate in the eikonal limit, but a simple double copy relation as in the massless 4d case does not arise. Instead, we show that information beyond the eikonal limit is required to construct the correct massive double copy. Nevertheless, we can construct a simple double copy for the phase shift. To further understand the double copy relation of topologically massive theories in the high-energy limit, we take a look at the classical solutions generated by a highly energetic particle in Section \ref{sec:shock}. We show that a coordinate space Kerr-Schild double copy can be obtained for wave solutions when taking into account a special set of boundary conditions. In the process, we show how the choice of $i \epsilon$ prescription for obtaining the phase shift is related to the boundary conditions of the topologically massive field. Lastly, we conclude in Section \ref{sec:discussion} by discussing other possible double copy relations and future directions.

\section{Topologically Massive Theories with Matter Couplings} \label{sec:TMT}
In this section we briefly review the actions of Topologically Massive Yang-Mills (TMYM) and Topologically Massive Gravity (TMG) theory with a minimal coupling to matter. The action of TMYM with a source is,
	\be\label{eq:tpYM}
	S_{TMYM}=\int d^3x\Bigg(-\frac{1}{4}F^{a\mu\nu}F_{a\mu\nu}+\epsilon_{\mu\nu\rho}\frac{m}{12}\left(6 A^{a\mu }\partial^{\nu}A^{\rho}_{a}+g\sqrt{2}f_{abc}A^{a\mu}A^{b\nu}A^{c\rho}\right)+\frac{g}{\sqrt{2}} A^{\mu a}J_{\mu a}\Bigg),
	\ee
where $m$ is the mass of the gauge field and $g$ the coupling strength. The equations of motion can be easily obtained from \eqref{eq:tpYM} and read

\begin{equation}
		D_\mu F^{\mu\nu}+\frac{m}{2}\varepsilon^{\nu\rho \gamma}F_{\rho\gamma}=\frac{g}{\sqrt{2}} J^\nu, \label{TMYM}
	\end{equation}
where $D_{\mu}=\partial_{\mu}-\frac{i g}{\sqrt{2}}A_{\mu}$, $F_{\mu\nu}=F^{a}_{\mu\nu}T^{a}$, with $F^{a}_{\mu\nu}$ the Yang-Mills field strength and $T^{a}$ the generators of the gauge group. A large simplification occurs when we consider an ansatz for the gauge field of the form $A^{\mu \ a}=c^a A^\mu$ such that the equations of motion become linear and read
	\begin{equation}
		\partial_\mu F^{\mu\nu}+\frac{m}{2}\varepsilon^{\nu\rho \gamma}F_{\rho\gamma}=\frac{g}{\sqrt{2}} J^\nu \label{TMYM},
	\end{equation}
where $F^{\mu\nu}$ is the Maxwell field strength since we have linearized the theory. 

On the gravitational side, we use the conventions $\kappa^2=16\pi G$ and  $g_{\mu\nu}=\eta_{\mu\nu}+\kappa h_{\mu \nu}$. Therefore, the action of TMG is,
	\be\label{eq:tpmGR}
	S_{TMG}=\frac{1}{\kappa^2}\int d^3x\sqrt{-g}\left(-R-\frac{1}{2m}\epsilon^{\mu\nu\rho}\left(\Gamma^{\alpha}_{\mu\sigma}\partial_{\nu}\Gamma^{\sigma}_{\alpha\rho}+\frac{2}{3}\Gamma^{\alpha}_{\mu\sigma}\Gamma^{\sigma}_{\nu\beta}\Gamma^{\beta}_{\rho\alpha}\right)+\L_{Matter}\right) \ ,
	\ee
and the equations of motion are,
\begin{equation}
G_{\mu\nu}+C_{\mu\nu}/m=-\kappa^2 T_{\mu\nu}/2 \ ,
\end{equation}
where $G^{\mu\nu}=R^{\mu\nu}-\frac{1}{2}R g^{\mu\nu}$ is the Einstein tensor and $C^{\mu\nu}=\epsilon^{\mu\alpha\beta}\nabla_{\alpha}(R^{\nu}_{\beta}-\frac{1}{4}g^{\nu}_{\beta}R)$ the Cotton tensor. The equations of motion largely simplify if we consider a Kerr-Schild ansatz for the graviton field $h_{\mu\nu}=\phi k_{\mu}k_{\nu}$ for which the equation of motion is linear.
	
\subsection{2-2 Scattering of Matter}
In this subsection, we look at the scattering of minimally coupled massive scalars through a topologically massive mediator and analyze their double copy relation \footnote{In all the scattering amplitude calculations presented here, we work in Lorenz gauge for TMYM and in de Donder gauge for TMG.}. We take the mass of the scalars to be that of the topologically massive mediators \footnote{When the mass of the scalar, $m$, is not the same as the mass of the mediator, $M$, the double copy of $A_4$ can be written as $M_4^{DC}=M_4+\frac{P(s,t,u)}{stu}$, where $P$ is a polynomial. This $\frac{P(s,t,u)}{stu}$ term has massless poles their residues are proportional to $m^2-M^2$. If we require that $M_4^{DC}$ only has contributions from the exchange of a massive mediator and contact terms, we have to set $M=m$.}. We write the tree level 2-2 scattering amplitude of scalars in the adjoint representation coupled to TMYM as 
\begin{equation}\label{scalar tmym}
    A_{4}=g^{2} \sum_{i=1}^{3} \frac{c_{i} n_{i}}{s_{i}-m^{2}} \ ,
\end{equation}
where the kinematic factors are given by
\begin{align}\label{eq:ns scalar tmym}
n_s&=\frac{i}{2}\left(u-t\right)-2m\frac{\epsilon_{\mu\nu\rho}p_1^{\mu}p_2^{\nu}p_3^{\rho}}{s},\nn\\
n_t&=\frac{i}{2}\left(s-u\right)-2m\frac{\epsilon_{\mu\nu\rho}p_1^{\mu}p_2^{\nu}p_3^{\rho}}{t},\nn\\
n_u&=\frac{i}{2}\left(t-s\right)-2m\frac{\epsilon_{\mu\nu\rho}p_1^{\mu}p_2^{\nu}p_3^{\rho}}{u},
\end{align}
where $s=-(p_1+p_2)^2$, $t=-(p_1+p_3)^2$ and $u=-(p_1+p_4)^2$. Here, the coupling to TMYM is given by Eq.~\eqref{eq:tpYM} with $J^{\mu \ a}= f^{abc}\partial^\mu\phi_b \phi_c$. Similarly, the minimally coupled scalar scattering amplitude in TMG is given as
\be\label{eq:ns scalar tmg}
\begin{split}
M_4=&\Bigg(\frac{8 \epsilon_{\mu\nu\rho}p_1^{\mu}p_2^{\nu}p_3^{\rho} m \left(4 m^2-2 s-t\right)-32 i m^4 s+8 i m^2 \left(s^2+s t+t^2\right)+i t^3}{t \left(m^2-t\right)}+\\
&\frac{8 \epsilon_{\mu\nu\rho}p_1^{\mu}p_2^{\nu}p_3^{\rho} m \left(-4 m^2+s+2 t\right)+i \left(-32 m^4 t+8 m^2 \left(s^2+s t+t^2\right)+s^3\right)}{s \left(m^2-s\right)}-\\
&\frac{i \left(-8 i \epsilon_{\mu\nu\rho}p_1^{\mu}p_2^{\nu}p_3^{\rho} m (s-t)+192 m^6-112 m^4 (s+t)+4 m^2 \left(5 s^2+8 s t+5 t^2\right)-(s+t)^3\right)}{\left(-4 m^2+s+t\right) \left(-3 m^2+s+t\right)}\Bigg)\frac{\kappa^2}{16},
\end{split}
\ee
where we have used $s+t+u=4m^2$ to express $u$ in terms of $s$ and $t$. The double copy of \eqref{scalar tmym}, $M_4^{DC}$, differs from \eqref{eq:ns scalar tmg} by 
\be
M_4-M_4^{DC}=-i m^2 \kappa^2,
\ee
which means that we can match them by adding a contact term, $-\frac{\kappa^2m^2}{4!}\phi^4$, in the action of TMG with a minimally coupled scalar. This non-trivial realization of the double copy reduces to the trivial case when taking the high-energy (large $s$ and small $t$) limit. In such limit, the contact term contribution becomes subdominant since the scattering through the topologically massive graviton grows as $s^2$. In the rest of this paper we will explore in detail the double copy in the eikonal limit and leave the analysis of the double copy with more general matter for future studies.

\section{Double Copy in the Eikonal Limit}
\label{sec:eikonal}
The high-energy limit of scattering processes has been largely studied due to its connections to classical backgrounds, which was first explored in \cite{HOOFT198761}. Recently, the focus on the eikonal limit has increased given the ability of obtaining classical observables that describe the inspiral phase of the coalescence of compact binaries from the phase shift \cite{KoemansCollado:2019ggb,Cristofoli:2020uzm,DiVecchia:2019myk,Parra-Martinez:2020dzs,DiVecchia:2021bdo,Heissenberg:2021tzo,Bern:2020buy,Bern:2020gjj,Damgaard:2021ipf,AccettulliHuber:2020oou,DiVecchia:2020ymx,DiVecchia:2021ndb,Haddad:2021znf,delaCruz:2017zqr}. In this limit, it has been shown that a simple double copy relation arises in 4 dimensions \cite{Saotome:2012vy,Melville:2013qca,Luna:2016idw}. Since the eikonal amplitude includes information at all loop orders, a double copy relation for topologically massive eikonal amplitudes will be the first hint for an all orders double copy.\\

We proceed to analyse in detail the topologically massive double copy in the eikonal limit where we expect it to hold without requiring extra interactions on the gravitational side. We consider the 2-2 scattering of external scalar fields with the following kinematics,
	\begin{align}
		p_1^\mu&=\left({1\over 2p^{v}}\left({{ q\ }^2\over 4}+m^2\right),p^{v} ,{ q \over 2}\right)\, , & p_3^\mu&=\left({-1\over 2 p^{v}}\left({{ q\ }^2\over 4}+m^2\right),-p^{v},{ q \over 2}\right)\, , \\
		p_2^\mu&=\left(p^{u},{1\over 2 p^{u}}\left({{ q\ }^2\over 4}+m^2\right), -{ q\over 2 }\right)\, , & p_4^\mu &=\left(-p^{u},{-1\over 2 p^{u}}\left( {{ q\ }^2\over 4}+m^2\right),{ -q\over 2 }\right)\,.
	\end{align}
These momenta are on-shell, $p_1^2=p_2^2= p_3^2=p_4^2=-m^2$, and satisfy the momentum conservation condition $p_1^\mu+p_2^\mu+p_3^\mu+p_4^\mu=0$. Here we work in lightcone coordinates $(u,v,x^1)$,
	\be u={1\over \sqrt{2}}\left(x^0- x^1\right)\,,\quad v={1\over \sqrt{2}}\left(x^0+ x^1\right)\,  . \ee
The independent Mandelstam invariants are
	\bea &&s=-(p_1+p_2)^2=\frac{(4m^2+8p_{v}p_{u}+q^2)^2}{32p_{u}p_{v}}\\
	&&t=-(p_1+p_3)^2=-{ q\ }^2\, .\eea
In the eikonal limit, the momenta $p^{v}$ and $p^{u}$ are much larger than $q$ and $m$ and hence  $s\approx -u >> t$ and $s>>m^2$. \\

In the following, we compute the eikonal amplitude to all orders for TMG and TME. We focus on the Abelian case for simplicity since we expect that in the eikonal limit both Abelian and non-Abelian cases double copy to the same gravitational solution, as in both cases the same diagrams contribute to the eikonal scattering amplitude. Also, we know that eikonal amplitudes are related to classical shock wave solutions, which are solutions to both Abelian and non-Abelian theories.
	
\subsection{Eikonal resummation in TMG}

In 4d, it has been shown that the ladder and cross-ladder diagrams for massive particles of arbitrary spin, which are expected to dominate in the eikonal limit, re-sum in impact parameter space \cite{Hinterbichler:2017qyt}. The eikonal $2-2$ amplitude to all loop orders is given by
	\begin{equation} \label{eq:M eikonal2}
		i \mathcal{M}^{\text {eik }}(s, t)=2 s \int \mathrm{d}^{D-2} \vec{b} e^{i \vec{q} \cdot \vec{b}}\left(e^{i \delta(s, \vec{b})}-1\right) \ ,
	\end{equation}
	where the eikonal phase reads
	\begin{equation} \label{eq:delta eikonal2}
		\delta(s, \vec{b})=\frac{1}{2 s} \int \frac{\mathrm{d}^{D-2} \vec{q}}{(2 \pi)^{D-2}} e^{-i \vec{q} \cdot \vec{b}} M_{\text{tree}}(s,t=-(\vec{q})^2) \ ,
	\end{equation}
	with $M_{\text{tree}}(s,t=-(\vec{q})^2)$ the 4-point, tree level scattering amplitude given by the t-channel graph in the eikonal limit. Furthermore, this phase can be expressed in terms of the square of 3-point amplitudes by applying a BCFW-like shift. We prove the eikonal resummation for topologically massive theories in appendix \ref{ap:eikonal res}. For TMG $M_{\text{tree}}$ is given as
	\be
M_{\text{tree}}(s,t=-(q^y)^2)=\frac{ -i\kappa^2 s^2 m }{ 2(q^y)^2(q^y+i m)}. \label{eq:Mtree}
\ee

To compute the phase shift explicitly, we see that we need to regulate the following divergent integral,
	\be \label{eq:delta integral}
\delta= \frac{-i \kappa^2 s m}{4}\int  \frac{d q}{ 2\pi} \frac{ 1 }{ q^2(q+i m)} e^{-i b q} \ ,
\ee
that is, we need to choose some $i \epsilon$ prescription for integrating around the pole at $q=0$. This freedom corresponds to the freedom in choosing boundary conditions. Following \cite{Edelstein:2016nml} we shift $q\rightarrow q-i \epsilon$ and close the integration contour in the lower half plane when $b>0$ and in the upper half plane when $b<0$. This way the contour at infinity goes to zero and we can evaluate \eqref{eq:delta integral} as a contour integral. It picks the residue of the pole at $q=-i m$ when $b>0$, and the residue of the pole at $q=0$ when $b<0$. Therefore we can write
\be 
\delta= \frac{ \kappa^2 s m}{4} \left(- \text{Res}_{q=-im}\left( \frac{ 1 }{ q^2(q+i m)} e^{-i b q} \right) \theta(b) +  \text{Res}_{q=0}\left( \frac{ 1 }{ q^2(q+i m)} e^{-i b q} \right) \theta(-b)\right).
\ee
Evaluating the residues we get that the phase shift is given by
\be 
\delta= \frac{ \kappa^2 s }{4m} \left(  e^{-m b }  \theta(b) +  (1-mb)  \theta(-b)\right) \ . \label{eq:phaseshiftTMG}
\ee
Then the eikonal amplitude reads 
\begin{equation} \label{eq:Meik integral}
	i{\cal M}_{\text{eik}}=2s
	\int db e^{-i b q} \left(\text{exp}\left(\frac{i \kappa^2 s }{4m} \left(  e^{-m b }  \theta(b) +  (1-mb) \theta(-b)\right)\right)-1\right) \ .
\end{equation}
Note that this could be explicitly evaluated in terms of incomplete gamma functions as in \cite{Deser:1993wt}.
\subsection{Eikonal resummation in TME}
The sum of all loop diagrams for TMYM is complicated due to the different colour factors arising at each loop order. Since we are interested in shock wave solutions which are also solutions of the linearised theory here we will consider the eikonal amplitude in topological massive electrodynamics (TME) of two scalars of charge $Q$. From now on we slightly change the notation by absorbing the $1/\sqrt{2}$ factor into $Q$. In other words, the covariant derivative acting on the scalar is now $D\phi=(d-i g Q A)\phi$. The calculation of the TME eikonal amplitude is given in \ref{ap:eikonal res} and the expressions are very similar to the TMG case: 
\begin{equation}
	i{\cal A}_{\text{eik}}=2s
	\int db e^{-i b q} \left(e^{i \delta}-1\right), \label{eq:eikexpA}
\end{equation}
where the phase shift reads
\begin{equation}
	\delta= \frac{1}{2s}\int \frac{d q^y}{ 2\pi} A_{\text{tree}}(s,t=-(q^y)^2) e^{-i b q^y},
\end{equation}
and $A_{\text{tree}}$ is given as
\be
A_{\text{tree}}(s,t=-(q^y)^2)=\frac{2 sg^2Q^2}{ q^y(q^y+i m)} .
\ee
Evaluating this explicitly and choosing the same contour of integration as in the TMG case gives
\be 
\delta= -i \frac{g^2Q^2}{2}2 \left(- \text{Res}_{q=-im}\left( \frac{ 1 }{ q(q+i m)} e^{-i b q} \right) \theta(b) +  \text{Res}_{q=0}\left( \frac{ 1 }{ q(q+i m)} e^{-i b q} \right) \theta(-b)\right),
\ee
which finally leads to a compact expression for the TME phase shift
\be 
\delta= -\frac{g^2Q^2 }{m}\left(  e^{-m b}\theta(b) +    \theta(-b)\right) \ . \label{eq:phaseshiftTME}
\ee

\subsection{Double Copy of Eikonal Amplitudes and Phase Shift}
After showing that the exponentiation in the eikonal limit is a feature of TMG and TME amplitudes just like in the gravity and Yang-Mills case in 4d, we would like to understand if a simple double copy relation arises in this limit just like in 4d \cite{Saotome:2012vy}. To do so, it is useful to write the $n-1$ loop diagrams of the TME eikonal amplitude as
\begin{equation}
\frac{{i\cal A}_{n-1}}{(\sqrt{2}g)^{2n}}=\frac{i}{n!}\left(\frac{1}{2s}\right)^{n-1}
	\int db e^{-i b q}\left(\int  \frac{ d q^y}{ 2\pi} { s Q^2 (q^y-i m) \over q^y((q^{y})^2+m^2)}e^{-i b  q^y}\right)^{n} \,.
\end{equation}
comparing this with ,
\begin{equation}
\frac{{i\cal A}_{n-1}}{(\sqrt{2}g)^{2n}}\sim
	\int  {c N\over D} \, ,
\end{equation}
where $c$ is the colour factor, $N$ is the kinematic factor and $D$ is the product of all propagators (which also includes the factor of $(2s)^{n-1}$ which comes from propagators), we can identify $c=Q^{2n}$, $N=\left(s(1-i m/q^y)\right)^n$ and $D=(2s)^{n-1}((q^{y})^2+m^2)^n$.
Following the prescription of leaving propagators untouched and exchanging colour (in this case electric charge $Q$) for kinematics, we can now find the double copy by considering the replacement $Q^2\rightarrow s(1-i m/q^y)$ which leads to 
\begin{equation}
\frac{{i\cal M}^{\text{D.C.}}_{n-1}}{(i \kappa/2)^{2n}}=
\frac{1}{n!}\left(\frac{1}{2s}\right)^{n-1}
	\int db e^{-i b q} \left(\int  \frac{ d q^y}{ 2\pi}\frac{s^2(q^y-i m)}{ (q^y)^2(q^y+im)} e^{-i b  q^y}\right)^n.
\end{equation}
When comparing to the TMG result in Eq.~\eqref{eikTMG}, Eq.~\eqref{eq:Mtree} and Eq.~\eqref{eq:Meik}, we can see that there is a mismatch in the amplitudes. Naively, this could be interpreted as requiring new degrees of freedom on the double copy side. Nevertheless, we will show that this is not the case, and instead it is just an artifact of the massive double copy.\\

We now proceed to understand the origin of the mismatch between the double copy and TMG eikonal amplitude by looking at the tree level result in detail. We start by looking at the eikonal limit of the kinematic factors of the four-point scalar amplitude in topological massive Yang-Mills:
\be
n_s=n_u=-i\frac{s}{2}, \quad n_t=s\left(i\pm\frac{m }{\sqrt{-t}} \right),
\ee
where the $\pm$ sign comes from $\epsilon_{\mu\nu\rho}p_1^{\mu}p_2^{\nu}p_3^{\rho}=\pm \frac{1}{2} \sqrt{s t u}$. We see that in the Yang-Mills amplitude the $t$ channel dominates since the $s$ and $u$ channels are suppressed by $1/s$:
\be \label{eq:aphi4 eikonal}
A_{\phi4}\rightarrow  g^2\frac{c_t n_t}{t-m^2}.
\ee
However, when constructing the massive double copy we have a new term proportional to $(n_s+n_t+n_u)^2$ coming from requiring the CK duality. In this term all channels contribute equally:
\be
n_s+n_t+n_u=\pm\frac{m s}{\sqrt{-t}} .
\ee
Therefore, the double copy of this amplitude is not simply proportional to $n_t^2$:
\be
\begin{split}
-i\left(\frac{\kappa}{2}\right)^{-2}M_4&=\frac{n_s^2}{s-m^2}+\frac{n_t^2}{t-m^2}+\frac{n_u^2}{u-m^2}-\frac{(n_s+n_t+n_u)^2}{m^2}\rightarrow \frac{n_t^2}{t-m^2}-\frac{(n_s+n_t+n_u)^2}{m^2}\\&=-\frac{2m s^2(m\pm i\sqrt{-t})}{t(t-m^2)} \ ,
\end{split}
\ee
which correctly reproduces the TMG eikonal amplitude. This tells us that to correctly double copy the scattering amplitude in the eikonal limit we require information beyond the eikonal limit. Alternatively, one could further require that $|t|\ll m^2$ in which case a simple double copy relation arises if we take $Q^2\rightarrow\frac{ms}{\sqrt{-t}}$ and note that in this limit the propagators are given by $-m^2$ \cite{Gonzalez:2021bes}. Nevertheless, restricting to the large mass limit would lead to an incorrect computation of the phase shift as can be seen from the previous sections. \\

Note that despite this issue at the level of the scattering amplitudes, a double copy for the phase shift will arise in the same way as in the 4d Yang-Mills and gravity case. To see this, one should note that given our choice of boundary conditions, the phase shift is only physical for $y>0$. On this side of the shock wave, the phase shift scales as expected for a scattering through a massive mediator of spin J, that is, $\delta \sim s^{J-1}e^{-mb}$. Thus we see that
\begin{equation}
\frac{\delta^{\text{TME}}}{g^2}=\frac{Q^2}{m} e^{-m b} \quad \xrightarrow[Q^2\rightarrow s]{} \quad  \frac{\delta^{\text{TMG}}}{(\kappa/2)^2}=\frac{s}{m} e^{-m b} \ .
\end{equation}

\section{Double Copy of Classical Solutions}
\label{sec:shock}

In this section, we will relate the eikonal amplitudes computed above to classical field profiles for the graviton and the gauge field. We do so by interpreting the 4-point scalar amplitudes as the scattering of a scalar off a shock wave background, which in turn is generated by a point-particle with large momentum (the second scalar). Since it is possible to write the gravitational shock wave in Kerr-Schild coordinates, we will explore if we can construct a classical double copy for such solutions. For the standard massless Yang-Mills and Gravity cases, the double copy of shock waves has been explored in various contexts\cite{Saotome:2012vy,Cristofoli:2020hnk,Bahjat-Abbas:2020cyb,Pasterski:2020pdk}. \\

We proceed by looking at the Kerr-Schild double copy, single copy and zeroth copy ansatze and understanding the equations of motion that they satisfy. Given a metric of the form
\begin{equation}
    g_{\mu \nu}=\eta_{\mu \nu}+\kappa k_{\mu} k_{\nu} \phi \ ,
\end{equation}
where $\eta_{\mu \nu}$ is the Minkowski metric and $k^\mu$ is null and geodetic, the single copy is given by
\begin{equation}
A^{a \ \mu}=c^a A^\mu= c^a k^\mu \phi \ .
\end{equation}
To understand if this ansatz gives a solution to TMYM we look at the trace reversed TMG equations with one upper and one lower index
	\begin{equation}
		R^\mu_\nu+\frac{1}{m}C^\mu_\nu=-\frac{\kappa^2}{2}(T^\mu_\nu-T g^\mu_\nu) \ .
	\end{equation}
Contracting this equation with a Killing vector $V^\mu$ one finds
	\begin{align}
		{\nabla}_{\lambda} F^{\lambda \mu}+\frac{1}{m}\epsilon^{\mu \alpha \beta}\nabla_\alpha\nabla^\lambda F_{\lambda\beta}+\frac{V^{\nu}}{V^{\lambda} k_{\lambda}}\left(X_{\nu}^{\mu}+Y_{\nu}^{\mu}\right)=\frac{\kappa}{2}J^{\mu}  \ ,\\
		J^{\mu} \equiv\frac{2V^{\nu}}{V^{\rho} k_{\rho}}\left(T_{\nu}^{\mu}-\delta_{\nu}^{\mu}T-\frac{1}{2m}\epsilon^{\mu}_{\alpha\nu}\nabla^\alpha T\right) \ ,
	\end{align}
where $\nabla$ is the covariant derivative of $\eta$ and 
	\begin{align}
		X_{\nu}^{\mu} \equiv-\bar{\nabla}_{\nu}\left[A^{\mu}\left(\bar{\nabla}_{\lambda} k^{\lambda}+\frac{k^{\lambda} \bar{\nabla}_{\lambda} \phi}{\phi}\right)\right] \ , \\
		Y_{\nu}^{\mu} \equiv F^{\rho \mu} \bar{\nabla}_{\rho} k_{\nu}-\bar{\nabla}_{\rho}\left(A^{\rho} \bar{\nabla}^{\mu} k_{\nu}-A^{\mu} \bar{\nabla}_{\rho} k_{\nu}\right) \ .
	\end{align}
This equation for the single copy largely simplifies when we consider wave solutions. In such case, the source either vanishes or corresponds to a particle sourcing a shock wave so that the trace of the stress energy tensor vanishes. Furthermore, we can work with lightcone coordinates such that 
\begin{equation}
    \eta_{\mu\nu}dx^\mu dx^\nu=-2du dv+dy^2 \ .
\end{equation}
Meanwhile, the Kerr-Schild and Killing vector are given by
\begin{equation}
k_\mu dx^\mu=-du \ , \quad V_\mu dx^\mu=dv\ , \quad k\cdot V=1
\end{equation}
The single copy equation of motion (eom) now reads
\begin{align}
		{\nabla}_{\lambda} F^{\lambda \mu}+\frac{1}{m}\epsilon^{\mu \alpha \beta}\nabla_\alpha\nabla^\lambda F_{\lambda\beta}=g J^{\mu}=2 g V^{\nu}T_{\nu}^{\mu} \ , \label{eq:singlecopy}
\end{align}
where we have taken $\kappa/2\rightarrow g$. We note that the single copy does not automatically satisfy the linearized equation of motion of TMYM unless the covariant derivatives pull out factors of the mass and give
\begin{equation}
\varepsilon^{\mu\rho \gamma}k_{\gamma}\nabla_\rho\left(\frac{\nabla^2 \phi}{m^2}\right)=\varepsilon^{\mu\rho \gamma}k_{\gamma}\nabla_\rho\phi \ .
\end{equation}
This is satisfied as long as the zeroth copy, $\phi^{a\tilde{a}}=c^a c^{\tilde{a}} \phi$, satisfies the linearized massive biadjoint scalar equation of motion for a vacuum solution or away of a localized source. To see that this is a consistent requirement, we obtain the zeroth copy eom by contracting Eq.~\eqref{eq:singlecopy} with the Killing vector $V$ and find
\begin{equation}
	\nabla^2\phi+\frac{m \epsilon^{\mu\lambda\rho}V_\mu(\nabla_\lambda\phi)k_\rho}{k\cdot V}+k\cdot Z= g \frac{J\cdot V}{k\cdot V}\equiv j \ ,
\end{equation}
where 
\begin{equation}
	Z^{\nu} \equiv\left(V^{\rho} k_{\rho}\right) \bar{\nabla}_{\mu}\left(\phi \bar{\nabla}^{[\mu} k^{\nu]}-k^{\mu} \bar{\nabla}_{\nu} \phi\right)+m\epsilon^{\mu\lambda\rho}V_\mu(\nabla_\lambda k_\rho)\phi \ .
\end{equation}
Considering again the case of wave solutions, we find that the zeroth copy satisfies the following equation of motion
\begin{equation}
	\nabla^2\phi+m \epsilon^{\mu\lambda\rho}V_\mu(\nabla_\lambda\phi)k_\rho =j=2 V^{\nu}V^{\mu}T_{\nu}^{\mu}\ .
\end{equation}
Requiring consistency of the double copy restricts the zeroth copy to satisfy $\partial_y \phi=-m \phi$. Thus, the Kerr-Schild double copy for TMG waves fixes the zeroth copy to satisfy
\begin{equation}
    \phi=A e^{-m y} \ , \label{eq:KSzerothcopy}
\end{equation}
where $A$ is a constant. It is trivial to see that plane waves will satisfy the double copy relation. Hence, in the following we analyze in detail the more involved case of shock wave solutions.

	\subsection{Shock Waves}
Shock wave solutions are closely related to scattering amplitudes in the eikonal limit. A probe particle moving in a shock wave background will experience a time delay which can also be computed by considering the 2 to 2 scattering in the eikonal limit of such particle with the massless particle generating the shock wave. Understanding the double copy of shock wave solutions could give a hint of an all order double copy relation. Here, we will analyze in detail how to construct a double copy for these classical solutions. In TMG, these solutions have been previously studied in \cite{Deser:1993wt,Edelstein:2016nml,Deser:1992nk} where an important feature is highlighted, the need to choose boundary conditions to fully fix the metric. In the following, we construct the TMG, TME and biadjoint scalar shock waves by choosing a special case of boundary conditions that makes the double copy relation explicit. \\

We start by constructing the shock wave solution in TMG for a source 
\begin{equation}
    T^{\mu\nu}=E \delta(u)\delta(y) \delta^\mu_v\delta^\nu_v \ ,
\end{equation}
with energy $E$. The metric can be written in lightcone Kerr-Schild coordinates with the Kerr-Schild scalar given by 
\begin{equation}
	\label{eq:TMGshock}
 \phi=\delta(u)g(y) \ ,
\end{equation}
where $g$ satisfies
	\begin{equation}
		g'''(y)+m g''(y)=\kappa E m  \delta(y) \ . \label{eq:eomShocTMG}
	\end{equation}
The TMG shock wave is not fixed by requiring asymptotic flatness as in the GR case. Since flatness only requires $g''(y)=0$, given a solution $g_1$ of Eq.~\eqref{eq:eomShocTMG}, $g_2=g_1+c(u) G$ is an asymptotically flat shock wave as long as $G''=0$. So one could ask if there are certain boundary conditions that allow for a double copy relation in coordinate space. Since we would like to connect our classical solution to the eikonal amplitudes, we will choose our boundary conditions such that they are consistent with the phase shift calculation in the previous section. \\

The 2-2 amplitude in the eikonal limit can be reproduced by considering the propagation of a point particle in the shock wave background. Following \cite{Deser:1993wt}\footnote{Note that there is a minus sign in front of $q$ in \eqref{eq:eikonal amp qm} compared to \cite{Deser:1993wt}, since the shock wave geometry is sourced by particle 1 and the incoming test particle is particle 2. Therefore $p_y-k_y$ of \cite{Deser:1993wt} is equal to $p_2^y+p_4^y=-q$ in our convention.}, we change the coordinate $v$ to
\be
v\rightarrow v+\frac{\kappa}{2} \theta(u)g(y)
\ee
which changes $dv\rightarrow dv+\frac{\kappa}{2} \left(\delta(u)g(y) du+\theta(u)d(g(y))\right)$ so the metric is now
\be
ds^2=-2du dv+dy^2-\kappa \; \theta(u) \; du \; d(g(y))\ .
\ee
For $u<0$ the metric if Minkowski in $u,v,y$ coordinates but for $u>0$ it is Minkowski in $u,v+\frac{\kappa}{2} g(y),y$ coordinates. Therefore, we can write the wavefunction of the incoming particle with momentum $p$ (in $u<0$ region) as
\be
\psi_{in}=\frac{1}{(2\pi)^{3/2}} e^{i p\cdot x} \ ,
\ee
while for outgoing particle of momentum $p'$ (in $u>0$ region) it is
\be
\psi_{out}=\frac{1}{(2\pi)^{3/2}} e^{i p'\cdot x+\frac{1}{2} i {p'}_v \kappa g(y) }\ .
\ee
The scattering amplitude $M_{eik}$ defined as
\be \label{eq:M pp from psi}
\delta(p_v-{p'}_v)\delta(p_u-{p'}_u)M_{eik}^{\text{p.p.}}(q_y={p'}_y-p_y)=\int d^3 x \psi_{in}(x) \psi_{out}^*(x) \ ,
\ee
is equal to
\be \label{eq:eikonal amp qm}
M_{eik}^{\text{p.p.}}=\int \frac{dy}{2\pi} e^{i\left(-q y -\frac{1}{2}\kappa {p'}_v  g(y)  \right)},
\ee
where for our kinematics ${p'}-p=-p_2-p_4=q$ and $p_u=E$ so ${p'}_v=\frac{s}{2E}$. This matches the result in \eqref{eq:Meik integral} if
\be
g(y)=-\kappa\frac{ E }{m } \left(  e^{-m y }  \theta(y) +  (1-m y) \theta(-y)\right),
\ee
when taking into account the non-relativistic normalization and conventions:
\begin{equation}
  M_{eik}^{\text{p.p.}}=\delta(q)+\frac{{\cal M}_{\text{eik}}}{4\pi s} \ . \label{eq:nonrelNorm}
\end{equation}
We can see that this choice gives boundary conditions such that in one side ($y>0$) of the shock wave the metric is Cartesian, i.e., $\lim_{y\rightarrow \infty} h_{\mu\nu}=0$, while for $y<0$ it is flat, even if it is in non-Cartesian coordinates. \\

We now proceed to compute the shock wave for linearized TMYM, that is, TME in a similar manner. Consider a source $J^\mu=Q \delta(u)\delta(y) \delta^\mu_v$ and an ansatz for the shock-wave solution in TMYM of the form
	\begin{equation}
		A^{a}=-c^a  \delta(u)f(y) du \ .
	\end{equation}
Plugging this in the TMYM eom gives
	\begin{equation}
		f''(y)+m f'(y)=- g Q \delta(y) \ . \label{eq:tmymShock}
	\end{equation}
As in the gravitational case, the shock wave is not fully determined by requiring that the field strength vanishes at infinity. In this case, given a solution $f_1$ of Eq.~\eqref{eq:tmymShock}, $f_2=f_1+c(u) F$ is also a shock wave with asymptotically vanishing field strength as long as $f'=0$. This leaves us with the freedom of imposing stronger boundary conditions on the gauge field to fully fix it. We will proceed as in the gravitational case and fix this boundary condition by looking at the eikonal scattering amplitudes. We consider the scattering amplitude for the propagation of a point particle in the shock wave gauge background.
Similar to the TMG case, we first perform a gauge transformation on $A$
\be
A\rightarrow A+d(\theta(u)f(y))=\theta(u) f'(y) dy \ .
\ee
The wavefunction of a point particle with charge Q moving in an electromagnetic field, satisfying $D^{\mu}D_{\mu}\psi=0$, can be written as
\be
\psi=e^{i  gQ\int^x A_{\mu} dx^{\mu}+i p\cdot x} \ ,
\ee
where the integral can be taken over any path that ends at $x$. We choose the path so that it starts in the $u<0$ region. The wavefunction of the incoming particle with momentum $p$ (in the $u<0$ region) is then
\be
\psi_{in}=\frac{1}{(2\pi)^{3/2}} e^{i p\cdot x} \ ,
\ee
while for the outgoing particle with momentum ${p'}$ (in the $u>0$ region) is
\be
\psi_{out}=\frac{1}{(2\pi)^{3/2}} e^{igQ\int^y f'(y') dy'+i {p'}\cdot x}=\frac{1}{(2\pi)^{3/2}} e^{igQ f(y)+\text{const.}+i {p'}\cdot x} \ .
\ee
Additionally, we choose the path such that the constant of integration is zero. Then by \eqref{eq:M pp from psi} the point-particle scattering amplitude, $A_{eik}^{\text{p.p.}}$, is equal to
\be
A_{eik}^{\text{p.p.}}=\int \frac{dy}{2\pi} e^{-i qy-igQ f(y) }.
\ee
Matching this to the eikonal amplitude in Eq.~\eqref{eq:phaseshiftTME} and \eqref{eq:eikexpA}, and taking into account the non-relativistic normalization of the point particle amplitude in Eq.~\eqref{eq:nonrelNorm} we find that
\begin{equation} \label{eq: tmym shock wave Cartesian}
		f(y)=\frac{gQ}{m}\left(e^{-my}\theta(y)+\theta(-y)\right) \ .
\end{equation}
This choice corresponds to boundary conditions in which  the field strength is zero for $y<0$ and on the other side of the shock wave we have $\lim_{y\rightarrow \infty} A^{\mu}=0$. \\

Lastly, we look at the zeroth copy, $\phi^{a \tilde{a}}=c^a c^{\tilde{a}} S$, shock wave which is a solution of the linearized bi-adjoint scalar equations of motion:
\begin{equation}
(\nabla^2-m^2)S=-\lambda \delta(u)\delta(y) \ .
\end{equation}
The scalar field shock wave solution is
\begin{equation}
S= \frac{\lambda}{2 m} \left(e^{-m y}\theta(y)+e^{m y}\theta(-y)\right)\delta(u) \ .
\end{equation}
Note that for the scalar case, there is no analog of having the curvature, or field strength vanishing, or an extra freedom in the solution from choosing boundary conditions. In fact, in this case the field approaches $0$ at both $y=\pm\infty$. \\

Now, we can proceed to construct the double copy of the TMYM shock wave to understand if it corresponds to the TMG shock wave. We can immediately see that this construction is highly dependent on our choice of boundary conditions. If we simply look at the equations of motion, we would naively conclude that the double copy of the TMYM shock wave does not correspond to the TMG shock wave. Instead, it would suggest that the source on the gravitational side is given by $T_{uu}=\frac{E}{m} \delta(u)\partial_y\delta(y)$ with all other components being zero. Nevertheless, one should be careful since we need to choose the appropriate boundary conditions to completely fix the shock wave solutions. Considering the special choice used in the computations above, we can see that the Kerr-Schild double copy holds on the $y>0$ side of the shock wave. In this side of the shock wave, the condition for the Kerr-Schild zeroth copy, Eq.~\eqref{eq:KSzerothcopy}, is fulfilled and the double copy relation is satisfied when we consider the replacements:
\begin{equation}
    \frac{\kappa}{2}\longleftrightarrow g \longleftrightarrow 1\ , \quad 2E \longleftrightarrow Q \longleftrightarrow \lambda\ , \label{eq:KSreplacements}
\end{equation}
where the factor of 2 is standard in relating the Kerr-Schild sources as seen in Eq.~\eqref{eq:singlecopy}. On the other hand, the relation does not hold for $y<0$, but this should not cause alarm, since on that side of the shock wave the spacetime is flat and the field strength vanishes. Hence, the apparent mismatch is simply explained by the choice of boundary conditions on that side of the shock wave which obscures the double copy relation.
This conclusion is similar to the time delay computation presented in \cite{Edelstein:2016nml}. Naively, computing the time delay, $\Delta x^{-}= \delta(s, b)/\left|p^{-}\right|$, using the phase shift in Eq.~\eqref{eq:phaseshiftTMG} and \eqref{eq:phaseshiftTME} will give a non-zero result on the $y<0$ which is unphysical since in this side of the shock wave the space is flat (the field strength vanishes).

\subsection{Gyratons}
Now we consider a generalization of the shock wave metric by adding a classical spin to the source. In this subsection, we will construct such solutions for TMG, TMYM, and the biadjoint scalar. In gravitational settings, this type of solutions have been dubbed gyratons and their metric is
    \be \label{eq:gyraton metric}
		ds^2=-2du dv+dy^2+\kappa \phi(u,y) du^2 +2\kappa \alpha(u,y)du dy \ .
	\ee
The stress tensor is now given by 
		\be
		T_{\mu\nu}=\left(E k_{\mu}k_{\nu}+  \sigma k_{(\mu }\epsilon_{\nu)}^{\alpha \beta} k_{\alpha} \partial_{\beta}\right)\delta(u)\delta(y) \ ,
\ee	
where $E$ is the energy of the source and $\sigma$ its spin. Writing $\phi=g(y)\delta(u)$, the TMG equation of motion now gives
	\begin{equation} 
		\partial_y^2\left(g'(y) \delta(u)-2\partial_u \alpha(u,y)\right)+m \partial_y\left(g'(y) \delta(u)-2\partial_u \alpha(u,y)\right)=\kappa  m \delta(u)  \left(E\delta(y) -\sigma \delta'(y) \right) \ . \label{eq:eomGyrTMG}
	\end{equation}
	We see that outside the sources the equation of motion is similar to that of the shock wave but now the $y$ derivative of $\phi$ is shifted to $\partial_y\phi=\partial_y\phi-2\partial_u \alpha$. The metric \eqref{eq:gyraton metric} is invariant under the following transformation \cite{Frolov:2005zq}:
	\begin{equation}
	    v\rightarrow v +\kappa \; \lambda(u,y), \quad \alpha\rightarrow \alpha-\partial_y  \lambda, \quad \phi\rightarrow \phi-2\partial_u \lambda \ .
	\end{equation}
	We can fix this gauge freedom by imposing
	\be
	\partial_y \alpha=0 \ .
	\ee
	 Then the solution of \eqref{eq:eomGyrTMG}, with the same boundary conditions for $\phi$ as before, is given as
	 \be
	 g=\frac{\kappa}{m}(E+m\sigma )e^{-my}\theta(y)+\frac{\kappa}{m}(E+m\sigma -E m y)\theta(-y) \ ,
	 \ee
	 \be
	 \alpha=0 \ .
	 \ee
Here we have chosen $\alpha=0$ so that the metric is in Cartesian coordinates on the $y>0$ side of the gyraton. Note that with this choice the metric is in Kerr-Schild coordinates as in the shock waves case.\\

It is interesting to note that the inclusion of classical spin changed the expression of the shockwave Kerr-Schild scalar on the physical side ($y>0$) by shifting the energy as 
\begin{equation}
    E \rightarrow E\left(1+m\frac{\sigma}{E} \right) \ .
\end{equation}
This type of energy shift was originally found when looking at gravitational anyons in \cite{PhysRevLett.64.611}. It is not surprising that the same shift arises for gyratons, since we can think of them as being sourced by highly-boosted anyons. Alternatively, this shift can be obtained by shifting the $y$ coordinate as $y\rightarrow y -\sigma /E$ and taking the small $\sigma / E$ limit. This shift is reminiscent of the spin deformations of 3-point on-shell amplitudes in 3d \cite{Burger:2021wss}, which in 4d are related to the Newman-Janis shift\cite{Guevara:2018wpp,Guevara:2019fsj,Bautista:2019tdr,Arkani-Hamed:2019ymq,Emond:2020lwi,Moynihan:2019bor}. We will see in the following that this shift also arises for the TME and biadjoint scalar gyratons.\\
	
On the gauge theory side we can consider the following gauge field:
	 \be
	 A^a=c^a ( \varphi(u,y) du+\beta(u,y) dy),
	 \ee
	 which gives only one non-vanishing component of field strength $F_{uy}=-\partial_y \varphi+\partial_u \beta$ just like in the shock wave case. Expressing $\varphi=f(y)\delta(u)$, the equation of motion with the spinning source, 
	 \be 
	 J_{\mu}=\left(Q k_{\mu}+  Q' \epsilon_{\mu}^{\alpha \beta} k_{\alpha} \partial_{\beta}\right)\delta(u)\delta(y) \ ,
	 \ee
	 gives the following:
	 \be
	 	\partial_y \left(f'(y)\delta(u)-\partial_u \beta(u,y)\right) +m  \left(f'(y)\delta(u)-\partial_u \beta(u,y)\right)=g \left(Q \delta(u)\delta(y)-Q' \delta(u)\delta'(y)\right) \ .
	 \ee \label{eq:eomGyrTMYM}
	We now choose to impose the Lorenz gauge condition which implies
	 \be
	 \partial_y \beta=0 \ ,
	 \ee
 but we still have some residual freedom from choosing boundary conditions which we fix by picking the same boundary conditions as in the shock waves case, that is, that the field strength vanishes on one side of the shock wave and on the other side the gauge field vanishes asymptotically. With this choice, we get the following solution:
	 \be
	 	f= g\frac{Q+m Q'}{m}\left(\theta(y)e^{-my}\right)+g\frac{Q}{m}\theta(-y),
	 \ee
	  \be
	 	\beta=0.
	 \ee
An important feature of this choice is that the gauge field is null, as required by the Kerr-Schild single copy ansatz. \\

Finally, we construct the zeroth copy solution for a spinning source. The linearized equation of motion reads
\begin{equation}
(\nabla^2-m^2)S=-\left(\lambda \delta(u)\delta(y)-\lambda' \delta(u)\delta'(y)\right) \ ,
\end{equation}
and its solution is given by
\begin{equation}
S= \frac{1}{2 m} \left((\lambda+m\lambda')e^{-m y}\theta(y)+(\lambda-m\lambda')e^{m y}\theta(-y)\right)\delta(u) \ .
\end{equation}
Consequently, we see that Kerr-Schild double copy works in a similar way as before in the region $y>0$ where the curvature (field strength) is non-zero, with the replacements now given by 
\begin{equation}
    \frac{\kappa}{2}\longleftrightarrow g \longleftrightarrow 1\ , \quad 2(E+m \sigma ) \longleftrightarrow Q+mQ' \longleftrightarrow \lambda+m\lambda' \ . 
\end{equation}

\subsection{dRGT Shock Waves}
As a special case, we will analyze the massive double copy of shockwaves in $d\geq 4$. Although it is known that the double copy construction fails to reproduce dRGT massive gravity at 5-points due to the appearance of spurious poles, it would be interesting to understand if it is possible that the 4-point double copy holds beyond tree-level. A simple example that that can help us understand this consists of analyzing the classical shock wave solutions since this can be entirely reproduced from looking at the 2-2 eikonal scattering. In dRGT, the shock wave solutions for a stress tensor of the form $T^{\mu\nu}=E \delta(u)\delta(\vec{x}-\vec{x}_0) \delta^\mu_v\delta^\nu_v$ can be written in Kerr-Schild form as
	\begin{equation} \label{eq:kerr ch}
		\mathrm{d} s^{2}=-2 \mathrm{~d} u \mathrm{d} v+\mathrm{d} \vec{x}^{2}+\kappa \delta(u) F\left(\vec{x}\right) \mathrm{d} u^{2} \ , \quad (-\nabla^2+m^2)F\left(\vec{x}\right)=\kappa E \delta(\vec{x}-\vec{x}_0) \ .
	\end{equation}
In fact, this is a solution to the equations of motion for a massive graviton with an arbitrary potential \cite{Mohseni:2011vv}, even if such cases include ghosts. The Kerr-Schild vector and scalar are given by
	\begin{equation}
		k_\mu d x^\mu=- d u \ , \quad
		\phi=\delta(u) F\left(\vec{x}\right) \ .
	\end{equation}
Thus, the single copy is given by
	\begin{equation}
		A^{\mu \ a}=-c^a \delta(u) F\left(\vec{x}\right)  \delta^\mu_v \ . 
	\end{equation}
Using this in the massive Yang-Mills equations of motion and considering the replacements in Eq.~\eqref{eq:KSreplacements}, we find 
	\begin{equation}
		(-\nabla^2+m^2)F\left(\vec{x}\right)\delta(u)=g Q \delta(\vec{x}-\vec{x}_0)\ ,
	\end{equation}
which tells us that this indeed corresponds to a shock wave solution of massive Yang-Mills with a source $J^\mu=Q \delta(u)\delta(\vec{x}-\vec{x}_0) \delta^\mu_v$. One should note that this double copy relation holds for all $d\geq 4$. This simple relation might be a hint that the loop level massive double copy holds at 4-points for massive gravity.

\section{Discussion} \label{sec:discussion}
We have analyzed the high-energy limit of topologically massive theories from two different perspectives. First, by looking at the scattering amplitudes in the eikonal limit; and second, by looking at the shock wave solutions for both a spinless and a spinning source. In the former analysis, we found that to construct the double copy of the eikonal amplitudes, we need information outside of the eikonal limit at tree-level. This is in stark difference with the massless $d\geq4$ case where a simpler double copy relation arises. In the latter, we obtained a double copy relation which is only manifest for a specific choice of boundary conditions. Along the way, we showed how the eikonal amplitude is related to the classical shockwave solutions and how the choice of $i \epsilon$ prescription required to regulate the phase shift corresponds to the choice of boundary conditions of the topologically massive field. This allowed us to choose the appropriate prescriptions to make the coordinate space double copy clear on the non-trivial (where the curvature and field strength are non-zero) side of the shockwave. This suggests that a cleaner double copy relation might arise when looking at the curvature and field strength, as in the $4d$ Weyl double copy, instead of looking directly at the fields, as in the Kerr-Schild double copy case. An immediate roadblock for finding an analogue of the Weyl double copy for topologically massive theories is the fact that the Weyl tensor vanishes in 3d. Instead, one can look at the Cotton tensor which has similar properties to the Weyl tensor. This will be explored in \cite{Gonzalez:2022otg}. \\

Several open questions remain when it comes to fully understanding the massive double copy. Regarding scattering amplitudes, it has not been shown if the double copy relation holds for six and higher-point amplitudes or for loop corrections. In the special case of topologically massive theories, a complete understanding of the situation when including couplings to generic matter is lacking. Some progress has been made in \cite{Moynihan:2020ejh,Burger:2021wss,Moynihan:2021rwh} and we have contributed to clarifying the situation in the high-energy limit in this paper. Nevertheless, a broader exploration for more generic sources for both classical solutions and scattering amplitudes is still missing.

\section{Acknowledgments}
We would like to thank Andrew J. Tolley for useful comments. MCG would like to thank Nathan Moynihan for useful discussions. MCG is supported by the European Union’s Horizon 2020 Research Council grant 724659 MassiveCosmo ERC–2016–COG and the STFC grants ST/P000762/1 and ST/T000791/1. JR is supported by an STFC studentship. AM is partially supported by the \emph{Séjours Scientifiques de Haut Niveau} fellowship. AM thanks \emph{L'Institut de Physique Théorique} and \emph{L'Institut des Hautes Études Scientifiques} for  their hospitality during the completion of this work.

	\pagebreak
	
\appendix

\section{Derivation of Eikonal Resummation} \label{ap:eikonal res}
\begin{figure}[H]
	\centering
		\includegraphics[width=5cm]{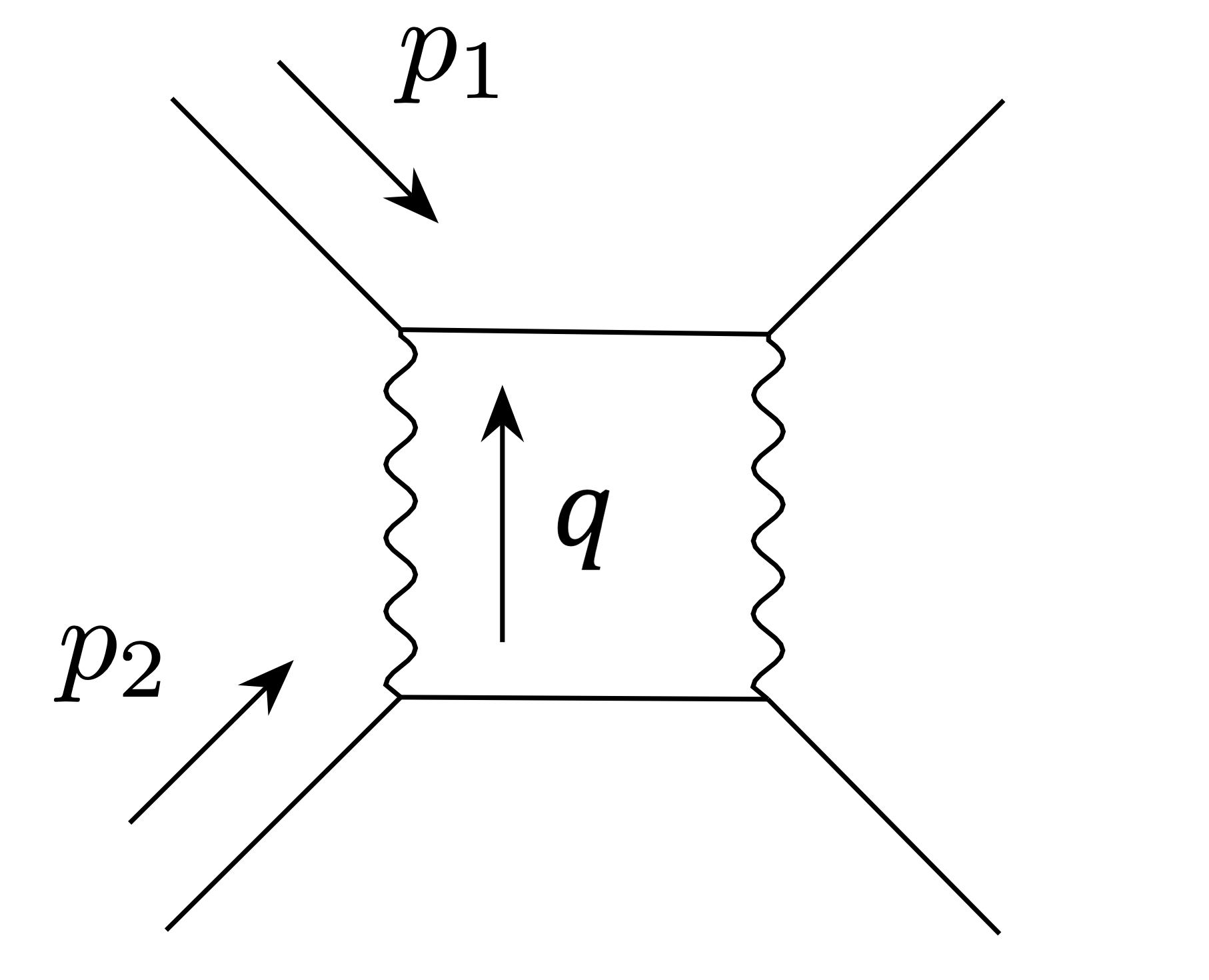}
	\caption{Example of a box diagram that appears within the ladder diagrams contributing in the eikonal limit. Here, the gravitons correspond to the rungs of the ladder.}
	\label{box}
\end{figure}

Here we will show that \eqref{eq:M eikonal2} and \eqref{eq:delta eikonal2} are valid in topological massive gravity by following the same steps as in \cite{Hinterbichler:2017qyt}. Assuming that only ladder diagrams contribute, the $n-1$ loop integrands are obtained by multiplying $n$ factors of two graviton-scalar-scalar vertices, contracted with a graviton propagator, together with scalar propagators, see figure \ref{box}.

Two scalar-scalar-graviton vertices contracted with a graviton propagator of momentum $q^{\mu}=(q^{u},q^{v},q^y)$ in the eikonal limit give
\begin{equation} \label{eq: excahnge diagram eikonal}
	-\frac{s^2}{4} \frac{F(q)}{q^4(q^2+m^2)},
\end{equation}
where $m$ is graviton mass and
\begin{equation}\label{eq:F}
	F(q)=\kappa^2\left(-i (q^{u} q^{v})^2+2 q^{u} q^{v} q^y m - 2 (q^y)^2(q^y-i m)m\right).
\end{equation}
Assuming that the momentum in the scalar propagators can be approximated as $(p+k)^2\rightarrow2p\cdot k$, where $p$ is $p_1$ or $p_2$ and $k$ is any loop momentum, the sum of $n-1$ loop diagrams gives the following:
\begin{align}
	i{\cal M}_{n-1}=&~
	\int \prod^n_{i=1} \left(\frac{d^3 q_i}{ (2\pi)^3} { -s^2 F(q_i)\over 4 q_i^4(q_i^2+m^2)}\right) \,(2\pi)^3\delta^3\left(p_3+p_1+\sum_{i=i}^n q_i\right) \nn\\
		&\times  {-i\over 2p_1\cdot q_1-i\epsilon}{-i\over 2p_1\cdot (q_1+q_2)-i\epsilon}\cdots {-i\over 2p_1\cdot (q_1+q_2+\cdots+q_{n-1})-i\epsilon} \nn\\
		& \times \sum_{\sigma \in S_n} {-i\over -2p_2\cdot q_{\sigma(1)}-i\epsilon}{-i\over -2p_2\cdot (q_{\sigma(1)}+q_{\sigma(2)})-i\epsilon}\cdots {-i\over -2p_2\cdot (q_{\sigma(1)}+q_{\sigma(2)}+\cdots+q_{\sigma(n-1)})-i\epsilon}\, .
	\end{align}
Using light-cone coordinates and taking the eikonal approximation we can write 
$p_1\cdot q=-p^{v} q^{u}$ and $p_2\cdot q=-p^{u} q^{v}$; hence
\begin{align} \label{eq: Meik in coord}
	i{\cal M}_{n-1}=&\frac{1}{(4p^{v} p^{u})^{n-1}}
	\int \prod^n_{i=1} \left(\frac{d q_i^{v}d q_i^{u} d q_i^y}{ (2\pi)^3}{ -s^2 F(q_i)\over 4 q_i^4(q_i^2+m^2)}\right) \,(2\pi)^3\delta\left(q+\sum_{i=i}^n q_i^y\right) \nn\\
	&\times \delta\left(\sum_{i=i}^n q_i^{u}\right)  {-i\over  -q_1^{u}-i\epsilon}{-i\over  -(q_1^{u}+q_2^{u})-i\epsilon}\cdots {-i\over  -(q_1^{u}+q_2^{u}+\cdots+q_{n-1}^{u})-i\epsilon} \nn\\
	& \times \delta\left(\sum_{i=i}^n q_i^{v}\right) \sum_{\sigma \in S_n} {-i\over  q_{\sigma(1)}^{v}-i\epsilon}{-i\over  (q_{\sigma(1)}^{v}+q_{\sigma(2)}^{v})-i\epsilon}\cdots {-i\over  (q_{\sigma(1)}^{v}+q_{\sigma(2)}^{v}+\cdots+q_{\sigma(n-1)}^{v})-i\epsilon}\, .
\end{align}
We now make use of the following identity,
\be
\begin{split}
	&\lim_{\epsilon\rightarrow 0}\delta(x_1+x_2+\cdots+x_n)\sum_{{\rm \sigma \in S_n}}{1\over x_{\sigma(1)}\pm i\epsilon}{1\over x_{\sigma(1)}+x_{\sigma(2)}\pm i\epsilon}\cdots {1\over x_{\sigma(1)}+x_{\sigma(2)}+\cdots x_{\sigma(n-1)}\pm i\epsilon}\\ 
	&=(\mp 2\pi i)^{n-1}\delta(x_1)\delta(x_2)\cdots \delta(x_n)\,, \label{eq:identity}
\end{split}
\ee
on the last line of \eqref{eq: Meik in coord} to get
	\begin{align} 
	i{\cal M}_{n-1}=&\left(\frac{2\pi i}{4p^{v} p^{u}}\right)^{n-1}
	\int \prod^n_{i=1} \left(\frac{d q_i^{v}d q_i^{u} d q_i^y}{ (2\pi)^3} { -s^2 F(q_i)\over 4 q_i^4(q_i^2+m^2)}\right) \,(2\pi)^3\delta\left(q+\sum_{i=i}^n q_i^y\right) \nn\\
	&\times \delta\left(\sum_{i=i}^n q_i^{u}\right)  {1\over  q_1^{u}+i\epsilon}{1\over  (q_1^{u}+q_2^{u})+i\epsilon}\cdots {1\over  (q_1^{u}+q_2^{u}+\cdots+q_{n-1}^{u})+i\epsilon} \nn\\
	& \times \prod_{i=1}^n \delta\left(q_i^{v}\right) \, .
\end{align}
Performing all $q_i^{v}$ integrals sets all $q_i^{v}$ to zero, so $q_i^2=-2q_i^{v}q_i^{u} + (q_i^y)^2\rightarrow (q_i^y)^2$, and we get
	\begin{align} 
	i{\cal M}_{n-1}=&\left(\frac{2\pi i}{4p^{v} p^{u}}\right)^{n-1}
	\int \prod^n_{i=1} \left(\frac{d q_i^{u} d q_i^y}{ (2\pi)^3} { -s^2 F(\{q_i^{u},0,q_i^y\})\over 4(q_i^y)^4((q_i^y)^2+m^2)}\right) \,(2\pi)^3\delta\left(q+\sum_{i=i}^n q_i^y\right) \nn\\
	&\times \delta\left(\sum_{i=i}^n q_i^{u}\right)  {1\over  q_1^{u}+i\epsilon}{1\over  (q_1^{u}+q_2^{u})+i\epsilon}\cdots {1\over  (q_1^{u}+q_2^{u}+\cdots+q_{n-1}^{u})+i\epsilon}  \, .
\end{align}
We can symmetrize the second line by summing over all permutations of labels and dividing by $n!$ to get 
	\begin{align} 
	i{\cal M}_{n-1}=&\frac{1}{n!}\left(\frac{2\pi i}{4p^{v} p^{u}}\right)^{n-1}
	\int \prod^n_{i=1} \left(\frac{d q_i^{u} d q_i^y}{ (2\pi)^3} { -s^2 F(\{q_i^{u},0,q_i^y\})\over 4(q_i^y)^4((q_i^y)^2+m^2)}\right) \,(2\pi)^3\delta\left(q+\sum_{i=i}^n q_i^y\right) \nn\\
	&\times \delta\left(\sum_{i=i}^n q_i^{u}\right) \sum_{\sigma \in S_n}  {1\over  q_1^{u}+i\epsilon}{1\over  (q_1^{u}+q_2^{u})+i\epsilon}\cdots {1\over  (q_1^{u}+q_2^{u}+\cdots+q_{n-1}^{u})+i\epsilon}  \, .
\end{align}
Then applying the identity \eqref{eq:identity} on the last line we get
\begin{align} 
	i{\cal M}_{n-1}=&\frac{1}{n!}\left(\frac{2\pi i}{4p^{v} p^{u}}\right)^{n-1}
	\int \prod^n_{i=1} \left(\frac{d q_i^{u} d q_i^y}{ (2\pi)^3} { -s^2 F(\{q_i^{u},0,q_i^y\})\over 4(q_i^y)^4((q_i^y)^2+m^2)}\right) \,(2\pi)^3\delta\left(q+\sum_{i=i}^n q_i^y\right) \nn\\
	&\times (-2 \pi i)^{n-1} \prod_{i=1}^n \delta\left(q_i^{u}\right)  \, .
\end{align}
Now performing $q_i^{u}$ integrals gives
\begin{equation} 
	i{\cal M}_{n-1}=\frac{1}{n!}\left(\frac{(2\pi)^2}{4p^{v} p^{u}}\right)^{n-1}
	\int \prod^n_{i=1} \left(\frac{ d q_i^y}{ (2\pi)^3} { -s^2 F(\{0,0,q_i^y\})\over 4(q_i^y)^4((q_i^y)^2+m^2)}\right) \,(2\pi)^3\delta\left(q+\sum_{i=i}^n q_i^y\right)   \, .
\end{equation}
From \eqref{eq:F} we get $F(\{0,0,q_i^y\})=- 2\kappa^2 (q^y)^2(q^y-i m)m$, and we can write \be
\delta\left(q+\sum_{i=i}^n q_i^y\right)=\int \frac{db}{2 \pi} e^{-i b\left(q+\sum_{i=i}^n q_i^y\right)}
\ee
so finally we get
	\begin{equation}\label{eikTMG}
	i{\cal M}_{n-1}=\frac{1}{n!}\left(\frac{1}{2s}\right)^{n-1}
	\int db e^{-i b q} \left(\int  \frac{ d q^y}{ 2\pi} { \kappa^2 s^2 m \over 2(q^y)^2(q^y+i m)} e^{-i b  q^y}\right)^n \,   \, ,
\end{equation}
where we used the fact that $s=2p^{v} p^{u}$ in the eikonal limit. The term in the second integral can be written as
\be
\frac{ \kappa^2 s^2 m }{ 2(q^y)^2(q^y+i m)}= i M_{\text{tree}}(s,t=-(q^y)^2), \label{eq:Mtree}
\ee
where $M_{\text{tree}}$ is the eikonal limit of tree level 2-2 scalar scattering amplitude. Then summing all loop diagrams gives the full eikonal amplitude:
\begin{equation}
	i{\cal M}_{\text{eik}}=2s
	\int db e^{-i b q} \sum_{n=1}^{\infty} \frac{1}{n!} \left( \frac{i}{2s}\int  \frac{d q^y}{ 2\pi} M_{\text{tree}}(s,t=-(q^y)^2) e^{-i b q^y}\right)^n  =2s
	\int db e^{-i b q} \left(e^{i \delta}-1\right), \label{eq:Meik}
\end{equation}
where 
\begin{equation}
	\delta= \frac{1}{2s}\int  \frac{d q^y}{ 2\pi} M_{\text{tree}}(s,t=-(q^y)^2) e^{-i b q^y}\ .
\end{equation}
is commonly referred to as the phase shift. Therefore, we have proved that the TMG amplitudes exponentiate in the eikonal limit, which to the best of our knowledge has not been proven before.\\

The calculation of the eikonal amplitude in TME is almost identical to that of TMG but now two scalar-scalar-photon vertices contracted with photon propagator gives
\be
\frac{s F(q)}{q^2(q^2+m^2)},
\ee
where
\be
F(q)=2g^2Q^2(-i q^{u} q^{v} + q^y (i q^y + m))
\ee
instead of \eqref{eq: excahnge diagram eikonal} and \eqref{eq:F}. Now, repeating the same steps as before we get 
\begin{equation} 
	i{\cal A}_{n-1}=\frac{1}{n!}\left(\frac{(2\pi)^2}{4p^{v} p^{u}}\right)^{n-1}
	\int \prod^n_{i=1} \left(\frac{ d q_i^y}{ (2\pi)^3} { s F(\{0,0,q_i^y\})\over (q_i^y)^2((q_i^y)^2+m^2)}\right) \,(2\pi)^3\delta\left(q+\sum_{i=i}^n q_i^y\right)   \, .
\end{equation}
Using $F(\{0,0,q_i^y\})= i 2g^2Q^2 q^y(q^y-i m)$ and the expression for the tree-level scattering amplitude in the eikonal limit,
\be
\frac{2i sg^2Q^2}{ q^y(q^y+i m)}=i A_{\text{tree}}(s,t=-(q^y)^2) \ ,
\ee
we get the same expression as in TMG case
\begin{equation}
	i{\cal A}_{\text{eik}}=2s
	\int db e^{-i b q} \left(e^{i \delta}-1\right), \label{eq:eikexpA}
\end{equation}
where the phase shift reads
\begin{equation}
	\delta= \frac{1}{2s}\int \frac{d q^y}{ 2\pi} A_{\text{tree}}(s,t=-(q^y)^2) e^{-i b q^y}.
\end{equation}
\bibliographystyle{JHEP}
\bibliography{references}
	
\end{document}